\documentclass[preprint,preprintnumbers,amsmath,amssymb]{revtex4}
\usepackage{graphicx}% Include figure files
\usepackage{dcolumn}% Align table columns on decimal point
\usepackage{bm}% bold math

\begin{document}

\title{Dynamics of quantum entanglement in the reservoir with memory effects}
\author{Xiang Hao}
\altaffiliation{Corresponding author,Email:110523007@suda.edu.cn}

\author{Jinqiao Sha}

\author{Jian Sun}

\affiliation{Department of Physics, School of Mathematics and
Physics, Suzhou University of Science and Technology, Suzhou,
Jiangsu 215011, People's Republic of China}

\author{Shiqun Zhu}

\affiliation{School of Physical Science and Technology, Suzhou
University, Suzhou, Jiangsu 215006, People's Republic of China}

\begin{abstract}
The non-Markovian dynamics of quantum entanglement is studied by the
Shabani-Lidar master equation when one of entangled quantum systems
is coupled to a local reservoir with memory effects. The completely
positive reduced dynamical map can be constructed in the Kraus
representation. Quantum entanglement decays more slowly in the
non-Markovian environment. The decoherence time for quantum
entanglement can be markedly increased by the change of the memory
kernel. It is found out that the entanglement sudden death between
quantum systems and entanglement sudden birth between the system and
reservoir occur at different instants.



\end{abstract}

\maketitle

{\Large \bf{1. Introduction}}

The quantum entanglement in composite systems is a key issue in
quantum information theory \cite{Divin00,Nielsen00}. The resources
of quantum entanglement are contained in nonclassical states which
are useful for quantum information processing, like quantum
teleportation, cryptography, dense coding and quantum computation
\cite{Bennett93,Divin98,Horodecki09}. It is known that entanglement
can be efficiently quantified by some useful measurement including
the concurrence \cite{Wootters98}, the negativity \cite{Vidal02} and
relative entropy \cite{Bennett96,Vedral98}. Because of interactions
between open systems and surrounding environments, the decoherence
of quantum entanglement is usually inevitable \cite{Breuer02}.
Therefore, dynamical properties of quantum entanglement have
received much attention
\cite{Sun07,Yu10,Cao08,Feng09,Nori10,Siomau10,Szallas10,Fanchini10,Erbe,Wang11,Lig11,Liao10,Fu10,Shi11}.
In some works \cite{Bellomo07, Li102}, the dynamics of the two-qubit
entanglement have been investigated when the noninteracting qubits
are respectively coupled to two independent reservoirs. It was found
out that the sudden death and revival of the entanglement can happen
at some instants. The Markovian approximation is used for the study
of dynamics of open systems weakly coupled to reservoirs without
memory effects. The corresponding map is always completely positive
and trace-preserving. However, in realistic environments with memory
effects, time evolutions of quantum systems often obey the
non-Markovian dynamics \cite{Breuer02}. The memory effects can be
described by some phenomenological master equations with a
reasonable memory kernel \cite{Imamoglu,Royer,Shabani05,Vacchini10,
Maniscalco06}. It is an obstacle to deal with physical decoherence
of open systems analytically and numerically in non-Markovian
environments. Among these master equations, the Shabani-Lidar one
interpolates between the exact dynamics and Markovian one
\cite{Shabani05}. The desirable property of Shabani-Lidar master
equation is the preservation of the complete positivity of the
reduced dynamics. The requirement of complete positivity of dynamics
ensures that the physical states of reduced systems can be applied
to practical tasks of quantum information processing. From the point
of view of a definite physical process, it is of great value to
obtain a completely positive dynamical map in a non-Markovian bath
and investigate the dynamics of quantum entanglement.

In this paper, by the Shabani-Lidar master equation, we construct
the completely-positive dynamical map in order to analyze the
decoherence of a two-level quantum system coupled to a vacuum
reservoir with general exponential memory. In Sec. 2, the reduced
dynamical map is written in the Kraus representation. The dynamics
of quantum entanglement between the open qubit and an ancillary
system isolated from the local reservoir is obtained in Sec. 3. The
memory effects on distribution of quantum entanglement between the
systems and local bath are also taken into account. A simple
discussion concludes the paper.

\vskip1.0cm

{\Large \bf{2. The completely positive reduced dynamical map}}

\vskip0.2cm

The decoherence of quantum system is always unavoidable because of
the interactions with its local reservoir. To set up the reduced
dynamical map, we investigate the total Hamiltonian of the effective
two-level open system and the local environment,
\begin{equation}
\label{eq:1}H=H_{S}+H_{E}+H_{SE},
\end{equation}
where
\begin{equation}
\label{eq:2}H_{S}=\omega_{0}\sigma_{+}\sigma_{-}
\end{equation}
describes the intrinsic Hamiltonian of the open two-level system,
\begin{equation}
\label{eq:3}H_{E}=\sum_{j}\omega_{j}a^{\dag}_{j}a_{j}
\end{equation}
is the Hamiltonian of the reservoir with a large amount of
independent harmonic oscillators and
\begin{equation}
\label{eq:4}H_{SE}=\sum_{j}g_{j}(\sigma_{+}a_{j}+\sigma_{-}a^{\dag}_{j})
\end{equation}
represents the interaction between the open qubit $S$ and the
environment $E$. The parameter $\omega_{0}$ describes the transition
energy from the ground state $|g\rangle$ to the excited one
$|e\rangle$. The operators $\sigma_{\pm}=(\sigma_{x}\pm i
\sigma_{y})/2$ are the raising and lowing operators of the open
qubit. Here $a^{\dag}_j,a_j$ are the creation and annihilation
operators of the reservoir. In the environment with memory effects,
the time evolution of an effective two-level system can be
approximately obtained by the Shabani-Lidar master equation
\cite{Shabani05}. The master equation can give new insight into the
dynamics of non-Markovian quantum systems. The reduced states for
the open system $\rho_{S}(t)$ can be written as
\begin{equation}
\label{eq:5}\frac {\partial \rho_{S}(t)}{\partial t}=\mathcal{L}
\int_{0}^{t}f(t^{\prime})\exp(\mathcal{L}
t^{\prime})\rho_{S}(t-t^{\prime})dt^{\prime}.
\end{equation}
Here the rotating-wave approximation is adopted. In the following
analysis, the initial total state $\rho_{S}(0)\otimes |0\rangle_E
\langle 0|$ is assumed. $|0\rangle_E$ denotes the vacuum state of
the local reservoir. The fucntion $f(t^{\prime})$ is the
phenomenological kernel characterizing the memory effects from the
environment. The memory kernel function can be experimentally
determined by quantum state tomography \cite{Shabani05}. In order to
understand the physical meaning of the Shabani-Lidar memory kernel,
we stress that this master equation describes a situation in which a
measurement of the environment at time $t^{\prime}$ is followed by a
Markovian evolution, described in terms of continuous measurements
of the environment at times $t>t^{\prime}$ \cite{Shabani05,
Maniscalco06}. The time $t^{\prime}$ characterizes the bath memory
effects. The memory kernel is a function which assigns weights to
different measurements. The determination of the memory kernel is
related to the reservoir spectral density. The local Liouvillian
operator $\mathcal{L}$ can be expressed by
\begin{equation}
\label{eq:6}\mathcal{L}\rho_{S}=\gamma_{0}\displaystyle
[\sigma_{-}\rho_{S}\sigma_{+}-\frac
12(\sigma_{+}\sigma_{-}\rho_{S}+\rho_{S}\sigma_{+}\sigma_{-})].
\end{equation}
The parameter $\gamma_{0}=\gamma_{0}(g_{j},\omega_{j},\omega_{0})$
describes the decaying rate in the Markovian model. For the
dynamical map $\Phi(t)$ of the reduced system $S$, the density
matrix at any time $t$ can also be expressed by
$\rho_{S}(t)=\Phi(t)\rho_{S}(0)$. According to \cite{Shabani05}, the
definite expression of the dynamical map is given by
$\Phi(t)\rho_{S}(0)=\sum_{j=0}^{3}\xi_{j}(t)\mathrm{Tr}[L_{j}\rho_{S}(0)]R_{j}
$. Here $L_{j}$ and $R_{j}$ are the $j$-th left and right damping
basis of the Liouvillian operator \cite{Briegel93}. In this case,
$\xi_{j}(t)=\mathrm{Lap^{-1}}[\frac {\displaystyle 1}{\displaystyle
s-\lambda_{j}F(s-\lambda_{j})}]$ where $\{
\lambda_{j}=1,-\gamma_{0},-\gamma_{0}/2,-\gamma_{0}/2 \}$ are the
four eigenvalues of the Liouvillian operator and the Laplace
transformation $F(s)=\mathrm{Lap}\displaystyle [f(t^{\prime})]$.

To investigate the conditions of complete positivity of the
dynamics, the Choi matrix $\mathcal{P}$ need be constructed by the
elements $\mathcal{P}(i,j)=\Phi(t)|i \rangle \langle j|$
\cite{Choi}. In the context, the general exponential memory kernel
function is considered as $f(t)=A\exp(-\gamma t)$ where $A$ is the
amplitude of the kernel and $\gamma$ represents the memory decaying
rate. This kind of memory kernel can be physically modeled by
\cite{Vacchini10}. The definite expression of the matrix
$\mathcal{P}$ can be obtained by
\begin{equation}
\label{eq:7}\mathcal{P}=\left(\begin{array}{cccc}
            \xi(\gamma_{0},t)&0&0&\xi(\gamma_{0}/2,t)\\
            0&1-\xi(\gamma_{0},t)&0&0\\
            0&0&0&0\\
            \xi(\gamma_{0}/2,t)&0&0&1
            \end{array}\right),
\end{equation}
where the elements $\xi(x,t)$ can be written as $
\xi(x,t)=\exp[-\frac {(1+g)xt}{2}](\cosh x\tilde{\omega}t+\frac
{1+g}{2\tilde{\omega}}\sinh x\tilde{\omega}t)$. The parameter is
$\tilde{\omega}=|\sqrt{(1+g)^2-4\alpha}|/2$ where $\alpha=A/x$ and
$g=\gamma/x$. If $\alpha > (1+g)^2/4$, the above function $\xi(x,t)$
is obtained by substituting $\sinh[\cdot]$ and $\cosh[\cdot]$ with
$\sin[\cdot]$ and $\cos[\cdot]$.

According to the results of \cite{Shabani05}, the complete
positivity of the map is equivalent to the positivity of the matrix
$\mathcal{P}$ where all of the eigenvalues are non-negative. The
condition of complete positivity of the map can be analytically
given by the inequality of
\begin{equation}
\label{eq:8}\xi^{2}(\gamma_{0}/2,t) \leq \xi(\gamma_{0},t) \leq 1.
\end{equation}
The impacts of the memory kernel on the complete positivity of the
dynamics are numerically calculated in Fig. 1. When $A\leq \gamma$
or $\alpha \leq g$, the values of
$D=\xi(\gamma_{0},t)-\xi^{2}(\gamma_{0}/2,t)$ are always
non-negative in Fig. 1(a). Because the values of $\xi(x,t)$ are
monotonically decreasing in time, the inequality of
$\xi(x,\infty)=0\leq \xi(x,t) \leq \xi(x,0)=1$ is always satisfied.
This means that the corresponding map at any time is completely
positive in the conditions of $\alpha \leq g$. However, if the
amplitude of the memory kernel satisfies $\alpha > (1+g)^2/4$, the
difference $D$ oscillates between the positive values and negative
ones in Fig. 1(b). For the time interval $\gamma_0
t\in(\pi-\arctan\frac {2\tilde{\omega}}{1+g},2\pi-\arctan\frac
{2\tilde{\omega}}{1+g})$, the values of $\xi(\gamma_0,t)$ are
negative which violates the positivity of the map. The region of
$D<0$ represents the non-Markovian dynamical map which contradicts
the complete positivity. According to the results of
\cite{Maniscalco06}, during the time evolution which violates the
positivity condition, the density matrix looses the probablistic
interpretation. The reduced states in this regions are not physical.
Therefore, the small amplitude of the kernel function is useful for
setting up the completely-positive dynamical map. In the following
analysis, we consider the physical evolution of the reduced states
in the region of $\alpha \leq g$.

In the condition of the complete positivity, the reduced dynamical
map can be expressed in the Kraus representation,
\begin{equation}
\label{eq:9}\rho_{S}(t)=\sum_{k}\hat{M}_{k}\rho_{S}(0)\hat{M}_{k}^{\dag}.
\end{equation}
The Kraus operators $\{\hat{M}_{k} \}$ can be deduced by the
completely-positive master equation with memory kernel. By the
method of \cite{Shabani05}, the $i$th column of the Kraus operator
$\hat{M}_k$ is the $i$th segment of the $k$th eigenvector of the
completely-positive matrix $\mathcal{P}$. The Kraus operators are
written as
\begin{align}
\label{eq:10}\hat{M}_{0}&=\sqrt{\mu_{+}}(a_{+}|e\rangle_S \langle
e|+b_{+}|g\rangle_S \langle g|)&\nonumber \\
\hat{M}_{1}&=\sqrt{1-\xi(\gamma_0,t)}|g\rangle_S \langle e|&\nonumber \\
\hat{M}_{2}&=\sqrt{\mu_{-}}(a_{-}|e\rangle_S \langle
e|+b_{-}|g\rangle_S \langle g|),&
\end{align}
where the parameters $\mu_{\pm}=\{1+\xi(\gamma_0,t)\pm
\sqrt{[1-\xi(\gamma_0,t)]^2+4\xi^2(\gamma_0/2,t)} \}/2$,
$a_{\pm}/b_{\pm}=(\mu_{\pm}-1)/\xi(\gamma_0/2,t)$ and
$b_{\pm}=\xi(\gamma_0/2,t)/\sqrt{\xi^2(\gamma_0/2,t)+(\mu_{\pm}-1)^2}$.
For the Markovian reservoir without memory, the Kraus operators can
be simplified into $\hat{M}_0=\nu|e\rangle_S \langle e|+|g\rangle_S
\langle g|$ and $\hat{M}_1=\sqrt{1-\nu^2}|g\rangle_S \langle e|$
where $\nu(t)=\exp(-\gamma_0 t/2)$.

\vskip1.0cm

 {\Large \bf{3. Dynamics of quantum entanglement}}

 \vskip0.2cm

As is known, decoherence of quantum entanglement actually happens
when local measurements are applied in quantum information
processing \cite{Nielsen00,Breuer02}. A reasonable model is
considered as one open qubit $S$ entangled with an ancillary qubit
$A$ which is isolated from a local vacuum reservoir $E$
\cite{Plenio}. The decoherence process can also be obtained by the
map in terms of the total system-environment state
\cite{Davidovich,Leung}. The evolution of the states for the total
system can be obtained by
\begin{equation}
\label{eq:11}\rho(t)=U_{SE}(t)[\rho_{SA}(0)\otimes |0\rangle_E
\langle0|]U^{\dag}_{SE}(t),
\end{equation}
where $\rho_{SA}(0)$ is the initial entangled state. The unitary
operator $U_{SE}(t)$ can be expressed by the Kraus operators,
\begin{equation}
U_{SE}(t)|m\rangle_{S}\otimes|0\rangle_E=\sum_{k}\hat{M}_{k}|m\rangle_S\otimes
|k\rangle_E.
\end{equation}

To clearly describe the dynamics of quantum entanglement between the
system and local reservoir, we consider the negativity $N$
\cite{Vidal02} as one general measure for quantum entanglement.
Based on the separability principle, the negativity $N$ is
introduced by,
\begin{equation}
\label{eq:12} N(\rho)=\max \{0,-2\sum_i\epsilon_i \}
\end{equation}
where $\epsilon_i$ is the $i$th negative eigenvalue of the partial
transpose of the mixed state. For the separability of unentangled
states, the partial transpose matrix has nonnegative eigenvalues if
the corresponding state is unentangled.

When the maximally initial state $\rho_{SA}(0)=\frac
12(|ee\rangle+|gg\rangle)(\langle ee|+\langle gg|)$ is chosen, the
reduced state $\rho_{SE}(t)=\mathrm{Tr}_{A}[\rho(t)]$ at any time is
written in the Hilbert space spanned by $\{ |ee\rangle, |eg\rangle,
|ge\rangle, |gg\rangle\}$,
\begin{equation}
\label{eq:13}\rho_{SA}(t)=\left(\begin{array}{cccc}
            \frac 12\xi(\gamma_0,t)&0&0&\frac 12
\xi(\gamma_0/2,t)\\
            0&0&0&0\\
            0&0&\frac {1-\xi(\gamma_0,t)}{2}&0\\
            \frac 12
\xi(\gamma_0/2,t)&0&0&\frac 12
            \end{array}\right).
\end{equation}
The decaying of quantum entanglement $N_{SA}$ is shown in Fig. 2(a).
It is seen that the values of $N_{SA}$ are always decreased to zero.
The effects of memory kernel function on the entanglement are
considered. When the relative amplitude $\alpha$ of the memory
kernel $f(t)$ is decreased, the vanishing of the entanglement
$N_{SA}$ occurs more slowly. To clearly describe the decaying
process, we can define the decoherence time $\tau$ where the value
of the entanglement is declined to $1/e$. Fig. 2(b) clearly shows
that the decoherence time scale is largely decreased with the
increasing of the relative amplitude. We demonstrate that the
decaying of quantum entanglement is much slower in the reservoir
with the memory effects. This point is beneficial for the
implementation of quantum computation in the non-Markovian
environments.

It is also interesting to study the distribution of quantum
entanglement between the systems and local reservoir. After the
tracing of the ancillary system $A$, the reduced states between the
open system $S$ and the local environment $\rho_{SE}$ can be
expressed in the Hilbert space spanned by $\{ |e0\rangle,
|e1\rangle,|e2\rangle, |g0\rangle, |g1\rangle, |g2\rangle \}$,
\begin{equation}
\label{eq:14}\rho_{SE}(t)=\frac 12\left(\begin{array}{cccccc}
            u&0&w&0&q_{+}&0\\
            0&0&0&0&0&0\\
            w&0&\xi(\gamma_0,t)-u&0&q_{-}&0\\
            0&0&0&v&0&z\\
            q_{+}&0&q_{-}&0& 1-\xi(\gamma_0,t)&0\\
            0&0&0&z&0&1-v
            \end{array}\right),
\end{equation}
where the elements are obtained by $u=\mu_+a^{2}_{+}$,
$v=\mu_+b^{2}_{+}$, $w=\sqrt{\mu_{+}\mu_{-}}a_{+}a_{-}$,
$z=\sqrt{\mu_{+}\mu_{-}}b_{+}b_{-}$ and
$q_{\pm}=\sqrt{\mu_{\pm}}a_{\pm}\sqrt{1-\xi(\gamma_0,t)}$. The
negativity $N_{SE}$ is also calculated and plotted as functions of
the relative amplitude $\alpha$ and the time scale $\gamma_0 t$ in
Fig. 3(a). It is shown that the sudden birth of entanglement occurs
at some time in the decoherence process and then the values are
rapidly decreased to zero after enough time. With the decreasing of
$\alpha$, the time interval for the increasing of $N_{SE}$ is
enlarged. Meanwhile, we also investigate the dynamics of the reduced
states $\rho_{AE}$ after the tracing of the open system $S$,
\begin{equation}
\label{eq:15}\rho_{AE}(t)=\frac 12\left(\begin{array}{cccccc}
            u&0&w&0&0&0\\
            0&1-\xi(\gamma_0,t)&0&p_{+}&0&p_{-}\\
            w&0&\xi(\gamma_0,t)-u&0&0&0\\
            0&p_{+}&0&v&0&z\\
            0&0&0&0&0&0\\
            0&p_{-}&0&0&z&1-v
            \end{array}\right),
\end{equation}
where some elements are given by $p_{\pm}=\sqrt{\mu_{\pm}}a_{\pm}$.
By the numerical calculation, the values of the entanglement
$N_{AE}$ are always increased to the maximum after enough time in
Fig. 3(b). This means that the initial entanglement can be
completely transferred to that between the ancillary system and
local bath during the decoherence process. It is also found out that
the entanglement between the ancillary system and independent bath
is produced when the open system is coupled to the local reservoir.

\vskip1.0cm

{\Large \bf{4. Discussion}}

\vskip0.2cm

We investigate the decoherence model of the open two-level system
coupled to a local vacuum reservoir using the Shabani-Lidar master
equation with memory kernel. For the general exponential memory, the
small amplitude of the kernel function and the large decaying rate
are helpful to construct the completely-positive reduced map. To
clearly describe the dynamics of quantum entanglement, we construct
the efficient map in the Kraus representation. By means of the
negativity, the entanglement distribution between the system and
local environment is also studied. It is found out that the
decreasing of quantum entanglement can be delayed pronouncedly in
the non-Markovian reservoirs with memory effects. From the point of
view of practical quantum information processing, the phenomenon is
helpful for the understanding of the non-Markovian quantum
computation. During the decoherence process, the sudden birth of
entanglement between the open system and local bath happens at some
time. After enough time, the initial entanglement will be
transferred to that between the ancillary system and independent
environment.

\vskip1.0cm

{\Large \bf{Acknowledgements}}

The work was supported by the Research Program of Natural Science
for Colleges and Universities in Jiangsu Province Grant No.
09KJB140009 and the National Natural Science Foundation Grant No.
10904104.

\newpage

{\Large Figure caption}

Figure 1

(a). The difference $D$ characterizing the complete positivity of
the map is plotted as a function of $\alpha$ and $\gamma_0 t$ when
the parameters are chosen by $\alpha \leq g=0.5$; (b). The
difference $D$ is plotted when the parameters $\alpha \gg g=0.1$.

Figure 2

(a). The dynamics of the entanglement between the open system and
ancillary one $N_{SA}$ is plotted when $g=0.5$. The solid line
denotes the case of $\alpha=0.1$, the dashed one represents that of
$\alpha=0.3$ and the dotted line describes the case of Markovian
reservoirs; (b). The decoherence time scale $\gamma_0\tau$ is
numerically calculated with the change of the relative amplitude
$\alpha$ of the kernel function.

Figure 3

(a). The entanglement $N_{SE}$ between the open system and local
reservoir is plotted as functions of $\gamma_0 t$ and $\alpha$ when
$g=0.5$; (b) The entanglement $N_{AE}$ between the ancillary system
and reservoir is also plotted.

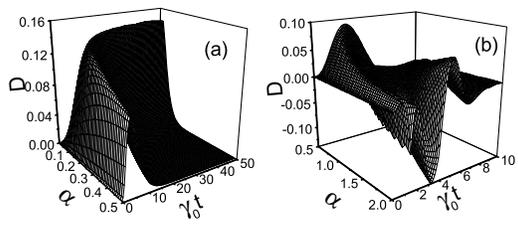
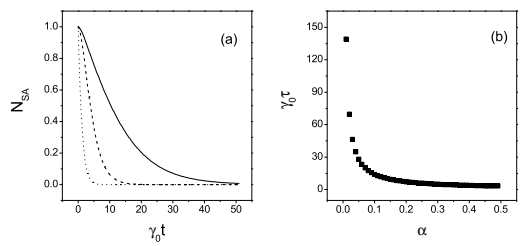
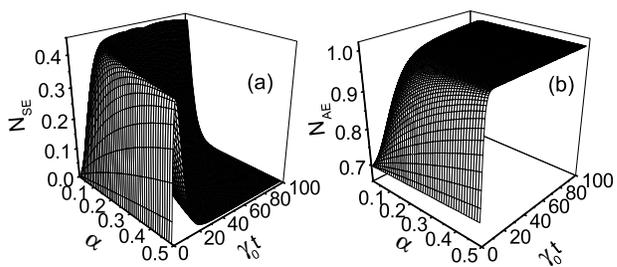
\end{document}